\documentclass[11pt]{article} \topmargin=-0.5cm \textheight 220mm
  \usepackage{amssymb}
\def\bline{\\ }\def\bbline{\\&& }\def\nbline{\\ \nonumber }\def\bbline{\\&& }\def\nbbline{\\ \nonumber &&}
\def\bline{ }\def\nbline{ }\def\bbline{ }\def\nbbline{ }
   \csname @addtoreset\endcsname{equation}{section}
    \csname @addtoreset\endcsname{figure}{section}
    \csname @addtoreset\endcsname{table}{section}

\newcounter{thanksnum}
\def\thanksnumber#1
{\setcounter{thanksnum}{\value{footnote}}\setcounter{footnote}{#1}%
                     \addtocounter{footnote}{-1}\footnotemark
                     \setcounter{footnote}{\value{thanksnum}}}
\def\newtheoremz#1{\@ifnextchar[{\@othmz{#1}}{\@nthmz{#1}}}

\def\@nthmz#1#2{%
\@ifnextchar[{\@xnthmz{#1}{#2}}{\@ynthmz{#1}{#2}}}

\def\@xnthmz#1#2[#3]{\expandafter\@ifdefinable\csname #1\endcsname
{\@definecounter{#1}\@addtoreset{#1}{#3}%
\expandafter\xdef\csname the#1\endcsname{\expandafter\noexpand
  \csname the#3\endcsname \@thmcountersepz \@thmcounterz{#1}}%
\global\@namedef{#1}{\@thmz{#1}{#2}}\global\@namedef{end#1}{\@endtheoremz}}}

\def\@ynthmz#1#2{\expandafter\@ifdefinable\csname #1\endcsname
{\@definecounter{#1}%
\expandafter\xdef\csname the#1\endcsname{\@thmcounterz{#1}}%
\global\@namedef{#1}{\@thm{#1}{#2}}\global\@namedef{end#1}{\@endtheoremz}}}

\def\@othmz#1[#2]#3{\expandafter\@ifdefinable\csname #1\endcsname
  {\global\@namedef{the#1}{\@nameuse{the#2}}%
\global\@namedef{#1}{\@thmz{#2}{#3}}%
\global\@namedef{end#1}{\@endtheoremz}}}

\def\@thmz#1#2{\refstepcounter
    {#1}\@ifnextchar[{\@ythmz{#1}{#2}}{\@xthmz{#1}{#2}}}

\def\@xthmz#1#2{\@begintheoremz{#2}{\csname the#1\endcsname}\ignorespaces}
\def\@ythmz#1#2[#3]{\@opargbegintheoremz{#2}{\csname
       the#1\endcsname}{#3}\ignorespaces}

\def\@thmcounterz#1{\noexpand\arabic{#1}}
\def\@thmcountersepz{.}
\def\@begintheoremz#1#2{ \trivlist \item[\hskip \labelsep{\bf #1\ #2}]}
\def\@opargbegintheoremz#1#2#3{ \trivlist
      \item[\hskip \labelsep{\bf #1\ #2\ (#3)}]}
\def\@endtheoremz{\endtrivlist}

\def\defi{\stackrel{{\scriptscriptstyle \Delta}}{=}}

\def\a{\alpha}

\def\O{\Omega}

\def\F{{\cal F}}
\def\w{\widehat}

\def\esssup{\mathop{\rm ess\, sup}}

\def\R{{\bf R}}
\def\E{{\bf E}}
\def\P{{\bf P}}

\def\s{\delta}

\def\ww{\widetilde}

\def\t{\theta}
\def\oo{\bar}
\def\s{\sigma}

\def\p{\partial}
\def\G{\Gamma}

\newcommand{\be}{\begin{equation}}
\newcommand{\ee}{\end{equation}}
\newcommand{\bd}{\begin{displaymath}}
\newcommand{\ed}{\end{displaymath}}
\newcommand{\ba}{\begin{array}{ll}}
\newcommand{\ea}{\end{array}}
\newcommand{\baa}{\begin{eqnarray}}
\newcommand{\eaa}{\end{eqnarray}}
\newcommand{\baaa}{\begin{eqnarray*}}
\newcommand{\eaaa}{\end{eqnarray*}}   

\def\ww{\tilde}

\def\u{ u }

\newtheorem{theorem}{Theorem}[section]

\title{Optimal replication of random claims by ordinary integrals with applications in
finance\footnote{This work  was supported by  ARC grant of Australia
DP120100928 to the author.}}
\author{ Nikolai Dokuchaev\footnote{Department of Mathematics \& Statistics, Curtin
University, GPO Box U1987, Perth,6845 Western Australia}}
\begin{document}
\maketitle
\begin{abstract} By the classical  Martingale Representation Theorem,
replication of random vectors can be achieved via stochastic
integrals or solutions of stochastic differential equations.  We
introduce a new approach to replication of random vectors via
adapted differentiable processes generated  by a controlled ordinary
differential equation. We found that the solution of this
replication problem exists and is not unique. This leads to a new
optimal control problem:  find a replicating process that is minimal
in an integral norm.  We found an explicit solution of this problem.
Possible applications to portfolio selection problems and to bond
pricing models  are suggested.
\end{abstract}
\section{Introduction}
By the classical  Martingale Representation Theorem,
 random variables generated by a Wiener process can be represented via
 stochastic integrals. This means that it is possible to find a Ito process
such that the terminal value matches a given random vector at a
fixed terminal time.
 This result leads to  the theory of
 backward stochastic differential equations and has many
 applications in Mathematical Finance.

We introduce a new approach to replication of random vectors via
adapted differentiable processes generated  by a controlled ordinary
differential equation. We found that the solution of this
replication problem exists and is not unique. Therefore, an optimal
control problem arises: to find a replicating control process that
is minimal in a certain norm.   We found an explicit solution of
this problem in linear quadratic setting. Possible applications to
portfolio selection problems and to bond pricing models  are
suggested.

\section{The control problem}
Consider a standard probability space $ (\Omega,\F, \P)$. Let $w(t)$
be a standard $d$-dimensional Wiener process which generates the
filtration $\F_t={\sigma\{w(r):0\leq r\leq t\}}$ augmented by all
the $\P$-null sets in $\F$; we assume that $w(0)=0$.

Let $f\in L_2(\O,\F_T,\P;\R^n)$ be a random vector.
\par
Let $a\in\R^n$, $A\in\R^{n\times n}$ be given, and let
$b\in\R^{n\times n}$ be a non-degenerate matrix. Let
$\G:[0,+\infty)\to \R^{n\times n}$ be a measurable matrix valued
function.
\par
Consider the following linear quadratic  optimal stochastic control
problem \baa&&\hbox{Minimize}\quad
\E\int_0^Tu(t)^\top\G(t) u(t)\,dt\quad\hbox{over}\quad u(\cdot)\label{optim4}\\
&&\hbox{ subject to}\nbbline\ba \frac{d x}{d t}(t) =A x(t)+ b
u(t),\quad t\in( 0,T)\\x(0)=a,\quad x(T)=f\quad
\hbox{a.s.}\ea\label{sys} \eaa
\par This is a stochastic control problem with  equality type constraints on the
value of the plant process that holds almost surely. This problem is
related to control problems for backward stochastic differential
equations (BSDEs) with fixed terminal value for the plant process.
In BSDEs control setting, a non-zero diffusion coefficient is
presented in the evolution equation for the plant process as an
auxiliary control process. The first problem of this kind was
introduced in \cite{DZ}.\index{Dokuchaev and Zhou (1999).} Our
setting is different: a non-zero diffusion coefficient is not
allowed. Similar problems were introduced in
\cite{D2010},\cite{D2013}.
\subsection{Admissible $\G$ and $u$}
For $p\ge 1$ and $q\ge 1$, we denote by $L_{p,q}^{n\times m}$ the
class of random processes $v(t)$ adapted to $\F_t$ with values in
$\R^{n\times m}$ such that
$\E\left(\int_0^T|v(t)|^qdt\right)^{p/q}<+\infty$. We denote by
$|\cdot|$ the Euclidean norm for vectors and the Frobenius norm for
matrices.
\par
 Let $g:[0,T)\to\R$ be a given
measurable function such that there exist $c>0$ and $\a\in (0.5,1)$
such that \baa \label{GG} && 0<g(t)\le c(T-t)^{\a},\quad\\
&&g(t)^{-1}\le c(1+(T-t)^{-\a}), \qquad t\in[0,T). \nonumber\eaa An
example of such a function is $g(t)=1$ for $t<T-T_1$,
$g(t)=(T-t)^\a$ for $t\ge T-T_1$, where $T_1\in(0,T]$ can be any
number.

Let $U$ be the set of all processes  from $L_{2,1}^{n\times 1}$ such
that  \baa \E\int_0^Tg(t)|u(t)|^2dt<+\infty. \label{u22}\eaa By the
definition of  $L_{2,1}^{n\times 1}$, it follows that, for $u\in U$,
\baa \E\left(\int_0^T|u(t)|dt\right)^2<+\infty . \label{u21}\eaa We
consider $U$ as the set of admissible controls.
\par
 We assume that  $\G(t)$ is a  measurable
matrix valued function in $\R^{n\times n}$, such that
$\G(t)=g(t)G(t)$, where $G(t)>0$ is a symmetric positively defined
matrix such that the matrices $G(t)$ and $G(t)^{-1}$ are both
bounded. By the definitions,
$\E\int_0^Tu(t)^\top\G(t)u(t)dt<+\infty$ for $u\in U$.

 Restrictions on the choice of $\G(t)=g(t)G(t)$ mean
that the penalty for the large size of $u(t)$ vanishes as $t\to T$.
Thus, we do not exclude fast growing $u(t)$ as $t\to T$ such that
$u(t)$ is not square integrable.
\subsection{Optimal control $u$}
 By the
Martingale Representation Theorem, there exists a unique $k_f\in
L_{2,2}^{n\times d}$ such that \baaa f=\E f+\int_0^{T}k_
f(t)dw(t).\label{Clarcfccc} \eaaa (See, e.g., Theorem 4.2.4 in
\cite{LL}, p.67).
\par
 We assume that there exists
$\tau\in(0,T)$ such that \baaa \esssup_{t\in[\tau,T]}\E
|k_f(t)|^2<+\infty.\label{kf}\eaaa
\par
Let \baaa \w k_\mu(t)=R(t)^{-1}k_f(t),\qquad R(s)\defi
\int_s^TQ(t)dt, \qquad\bline Q(t)= e^{A(T-t)} b\G(t)^{-1}b^\top
e^{A^\top(T-t)}. \eaaa By Lemma 1 from \cite{D2013}, it follows that
$\w k_\mu(\cdot)\in L_{2,2}^{n\times d}$.
\begin{theorem}\label{ThM}  Let \baaa \w\mu(t)=R(0)^{-1}(\E
f-e^{AT}a)+\int_0^t\w k_\mu(s)dw(s) \eaaa and \baaa \w
u(t)=\G(t)^{-1}b^\top e^{A^\top(T-t)}\w\mu(t).\eaaa Then this $u$
belongs to $U$ and it is a unique optimal solution  of problem
(\ref{optim4})-(\ref{sys}) in the class $u\in U$.
\end{theorem}\par
 In \cite{D2010}, a related problem  was considered for a simpler
case when it was required to ensure that $x(T)=\E\{f|\F_\t\}$ for
some $\t<T$.
 \section{Applications to finance}\label{SecF} Replication on the basis of
 Martingale Representation Theorem is the main tool in modern Mathematical Finance. The presented above approach to replication
 on the basis of Theorem \ref{ThM} can also be applied to problems arising in finance.
 Some possible applications are listed below,including optimal cash accumulation policy and modeling of the bond prices.
  \subsection{ Optimal cash accumulation policy}\label{sSec1}
 Consider a model where  a there is a risky asset with the price
$S(t)$ which is  a random continuous time process with positive
values adapted to a Wiener process $w(t)$. We assume that \baaa
dS(t)=S(t)[a(t)dt+\s(t)dw(t)],\eaaa where $a(t)$ and $\s(t)$ are
some $\F_t$-adapted bounded processes such that $\s(t)\ge C$ a.e.,
where $C>0$ is a constant. \par Let terminal time $T>0$ be given.
\par
Assume that an investor wishes to accumulate gradually an amount of
cash that allows  to purchase a share of this risky asset at time
$T$.  Let $u(t)$ be the process describing the density of the cash
deposits/withdravals at time $t\in(0,T)$, such that $u(t)\Delta t$
is the amount of cash deposited during the time interval
$(t,t+\Delta t)$, for a small $\Delta t>0$. Assume that it is
preferable that the cash flow will be as smooth as possible.
\par
Let us assume first  that the bank interest rate is zero, for both
loans and savings. In this case, the total amount of cash at the
terminal time will be $\int_0^Tu(t)dt$.

\par
Theorem \ref{ThM} can be applied now for $n=1$, $A=0$, $b=1$,
$f=S(T)$. In this case, \baaa  R(s)\defi \int_s^T\G(t)^{-1}dt. \eaaa
If $a(t)\equiv 0$ then Theorem \ref{ThM} ensures that the process
\baa\label{u3}
&&u(t)\nbbline=\G(t)^{-1}\Bigl[R(0)^{-1}S(0)+\int_0^tR(s)^{-1}dS(t)\Bigr]
\eaa is such that \baa \quad \int_0^Tu(t)dt=f\quad
\hbox{a.e.}\label{u12}\eaa Moreover,  this $u(t)$ is optimal in $U$
in the sense of the optimality criterion from Theorem \ref{ThM},
i.e., $\E\int_0^T\G(t)u(t)^2dt$ is minimal.

\par
Let us consider  a more general model where the bank interest rate
is $r\ge 0$, for both loans and savings. In this case, the total
amount of cash at the terminal time will be
$\int_0^Te^{r(T-t)}u(t)dt$.

\par
Theorem \ref{ThM} can be applied now for $n=1$, $A=r$, $b=1$,
$f=S(T)$. In this case, \baaa R(s)\defi \int_s^TQ(t)dt, \qquad Q(t)=
e^{2r(T-t)} \G(t)^{-1}. \eaaa If $a(t)\equiv 0$ then Theorem
\ref{ThM} ensures that the corresponding process (\ref{u3})  is such
that \baa \quad \int_0^Te^{r(T-t)}u(t)dt=f\quad
\hbox{a.e.}\label{u122}\eaa Again, this $u(t)$ is optimal in $U$ in
the sense of the optimality criterion from Theorem \ref{ThM}, i.e.,
$\E\int_0^T\G(t)u(t)^2dt$ is minimal.
\par
If $a(\cdot)\neq 0$, then conditions (\ref{u12}) are still satisfied
 for $u(t)$ defined by (\ref{u3}) but the value $\E\int_0^T\G(t)u(t)^2dt$ is not minimal over $u$ anymore.
Instead, $\E_Q\int_0^T\G(t)u(t)^2dt$ is minimal, where $\E_Q$ is the
expectation defined by an equivalent probability measure $Q$ such
that $S(t)$ is a martingale; we will call it a martingale measure.
This still means that deviations of $u$ are minimal but in a
different metric. It can be also noted that the definition of the
class $U$ for the original measure has to be adjusted for the new
measure $Q$, with the expectations $\E$ replaced by $\E_Q$.
\par
Let us consider a modification of the cash accumulating problem
where the accumulated cash amount has to be a given proportion of
the excess achieved by the equity at the terminal time. This problem
arises for a writer of a naked or a partially naked call option. To
cover this case, it suffices to apply Theorem \ref{ThM} with
$f=c\max(S(T)-K,0)$, where $c>0$ is the prescribed proportion. We
assume that $\s(t)$ is non-random and that the bank interest rate is
 $r>0$. In this case, it is well known that \baaa f=H(S(0),0)+\int_0^T\frac{\p H}{\p x}(S(t),t)dS(t). \eaaa
Therefore, $k_f(t)=\frac{\p H}{\p x}(S(t),t)\s(t)S(t)$ and
$\E_Qf=H(S(0),0)$. By Theorem \ref{ThM},
 (\ref{u122}) is ensured for this $f$ with \baaa
u(t)=c\G(t)^{-1}\Bigl[R(0)^{-1}H(S(0),0)\bline+\int_0^tR(s)^{-1}\frac{\p
H}{\p x}(S(t),t)dS(t)\Bigr], \label{u4}\eaaa where
$H(x,t)=\E_Q\left\{\max(S(T)-K,0)|S(t)=x\right\}$. This expectation
is under the martingale measure $Q$ again;  if $\s(t)$ is constant,
it can be calculated by the standard Black-Scholes formula.  In
addition, $\E_Q\int_0^T\G(t)u(t)^2dt$ is minimal among all $u$ such
that (\ref{u12}) is satisfied.

\par Another possible
choices of $f$ in this setting may include
 $f=c\min_{t\in[0,T]}S(t)$ (a given
proportion of the minimum achieved by the equity during the time
period $[0,T]$), or $f=\frac{c}{T}\int_0^TS(t)dt$ (a given
 proportion of the average equity value over the time period
$[0,T]$). Here $K>0$ and $c>0$ are some constants. In these cases,
$u(t)$ also can be represented explicitly for non-random $\s(t)$.

The model described above can also be applied to the  problems of
optimal dividend flow selection. In particular, it can be applied to
the setting where the manager of a firm with the capitalization
$S(t)$ wishes to pay dividends during the time period $[0,T]$ such
that the total payoff $\int_0^Tu(t)dt$ over this interval will be,
say $5\%$ of the equity $S(T)$ at time $T$.  The typical approach is
a barrier criterion of dividend payments or analysis of ruin times;
the methods are usually based on dynamic programming
(see,e.g.,\cite{DW},\cite{GS})\index{ Dickson and Waters (2004),
Gerber and Shiu (2004),} and the bibliography here). Theorem
\ref{ThM} leads to a new approach to this problem.
\subsection{Modelling of the bond prices} Consider continuous time bond
pricing model for zero coupon bonds. Let $B(t,T)$ be the bond price
at time $t$ for the zero coupon bond with payoff \$1 at time $T$,
where $T>t$. Let $r(t)$ be the short rate. We assume that the
process $r(t)$ is $\F_t$-adapted. Here $\F_t$ is the same as above;
it is the filtration generated by a Wiener process.
\par
We assume that the probability measure $\P$ is a measure used for
the pricing such that,  for a given process $r(t)$, \baaa
B(t,T)=\E\Bigl\{\exp\Bigl(
-\int_t^{T}r(s)ds\Bigr)\,\Bigr|\F_t\Bigr\}.\eaaa Let $\ww B(t,T)$ be
the discounted bond price defined as \baaa \ww B(t,T)=\exp\Bigl(
-\int_0^{t}r(s)ds\Bigr)B(t,T).\eaaa In particular, \baaa \ww
B(T,T)=\exp\Bigl( -\int_0^{T}r(s)ds\Bigr). \eaaa Usually, a model
evolution of $r(t)$ is being suggested first and then the
distributions of  $\ww B(t,T)$ and $B(t,T)$ are derived. Theorem
\ref{ThM} gives an alternative: to model first the distribution of
the random variable $\xi=\w B(T,T)$, and derive the evolution low of
$r(t)$ from the distribution of $\xi$. It appears that it can be
done for a quite wide class of random variables $\xi$ with the
values in $(0,1)$ such that $f=-\log \xi$ satisfies conditions of
Theorem \ref{ThM}. Since $\xi\in (0,1)$, we have that $f>0$ a.s.  By
Theorem \ref{ThM}, there exists an adapted process $r(t)$ such that
\baaa f= \int_0^{T}r(s)ds,\quad \xi=\exp\Bigl(
-\int_0^{T}r(s)ds\Bigr).\eaaa Moreover,  the process $r(t)$ can be
selected to be optimal meaning that it has minimal deviations (in
the sense of the optimality criterion  from Theorem \ref{ThM}).

This approach can be extended on the case of a bond market where
there are  bonds with different non-random maturity times $T_k$,
$k=1,...,N$, $T_k>T_{k+1}$ for all $k$. Assume that we are given
$\F_{T_k}$-measurable random variables $\xi_k$ with values in
$(0,1)$. Let $f_1=-\log \xi_1$, $f_2=\log \xi_1-\log \xi_2$,...,
$f_k=\log \xi_{k}-\log \xi_{k-1}$. Applying Theorem \ref{ThM}
modified for the positive initial times, we obtain that there exists
an adapted process $r(t)$ such that \baaa f_1= \int_0^{T_1}r(s)ds,\
\ldots,\ f_k= \int_{T_{k-1}}^{T_k}r(s)ds,\quad k=2,...,N,\eaaa and
\baaa \xi_k=\exp\Bigl( -\int_0^{T_k}r(s)ds\Bigr),\quad k=1,...,N.
\eaaa This leads to a bond market model such that \baaa \ww
B(T_k,T_k)=\xi_k,\quad k=1,...,N, \eaaa for an arbitrarily chosen
set $\{\xi_k\}$ such that
  the corresponding random variables $f_k$ has final second
moments and that condition (\ref{kf}) is satisfied for $T=T_k$ and
$f=f_k$. As we had mentioned already,  the conventional approach is
to select a model for the process $r(t)$ first and then to derive
$\ww B(T_k,T_k)$. This possibility to start with models for $\ww
B(T_k,T_k)$ is established here. This could give new opportunities
for modeling of bond prices.
\section{Proof of Theorem \ref{ThM}}   For the sake of
completeness, we give below the proof of Theorem \ref{ThM}; this
proof follows the proof of Theorem 1 from \cite{D2013}.
\par
Clearly, equation  (\ref{sys}) gives that  \baa x(t)=\int_0^t
e^{A(t-s)}bu(s)ds+e^{At}a.\label{solut}\eaa By (\ref{u21}), $x(T)\in
L_2(\O,\F_T,\P;\R^n )$ for any $u\in U$.\par
 Let the function $L(u,\mu):U\times L_2(\O,\F_T,\P;\R^n)\to\R$ be defined as
\baaa L(u,\mu)\defi\frac{1}{2}\E\int_0^Tu(t)^\top
\G(t)u(t)\,dt+\E\mu^\top(f-x(T)). \eaaa For a given $\mu$, consider
the following problem: \baa \hbox{Minimize $L(u,\mu)$ over $u\in
U$.}\label{min}\eaa This problem  does not have constraints on
terminal value $x(T)$. Therefore, it can be solved by usual
stochastic control methods for the forward plant equations.
 We solve problem (\ref{min}) using the so-called
 stochastic maximum principle that gives a necessary condition of optimality; see, e.g., \cite{Arkin}-\cite{CaKa},
  \cite{D1988}-\cite{DZ}, \cite{Haussmann}-\cite{Kushner}, \cite{Peng}-\cite{YZ}).
 \index{Arkin and Saksonov
  (1979), Bismut (1978),  Cadenillas and Karatzas (1995), Bensoussan (1983),
Dokuchaev (1988), Haussmann (1986),
 Kushner (1972), Peng (1990), Yong and Zhou (1999)).} For our problem ({\ref{min}),
 all versions of the stochastic maximum principle from the cited papers  are equivalent and can be formulated as the following:
 if $u=\u_\mu\in U$
 is optimal then
 \baa  \label{smp1}\psi(t)^\top bu_\mu(t)-\frac{1}{2}u_\mu(t)^\top \G(t)
 u_\mu(t)\nbline\ge \psi(t)^\top bv-\frac{1}{2}v^\top \G(t)v
\eaa for a.e. $t$ for all $v\in\R^n$ a.s., where $\psi(t)$ is a
process from $L_{2,2}^{n\times 1}$ such
that \baaa &&d\psi(t) =- A^\top \psi(t) dt+\chi(t)dw(t),\\
&&\psi(T)=\mu, \eaaa for some process $\chi\in L_{2,2}^{n\times n}$.
(See, e.g., Theorem 1.5 from \cite{CaKa},
\index{Cadenillas and Karatzas (1995),} p.609).
The only solution of the backward equation for $\psi$ is \baa
\psi(t)=e^{A^\top(T-t)}\mu(t),\quad \mu(t)=\E\{\mu|\F_t\}.
\label{psi}\eaa
\par
Necessary conditions of
 optimality (\ref{smp1}) are satisfied for a unique up to equivalency process  $u= u_\mu$
defined as \baa &&
u_{\mu}(t)=\G(t)^{-1}b^\top \psi(t). \label{SMP}\eaa
 Let us show that $u_\mu\in
U$ for any $\mu$. We have that \baaa\E\left(\int_0^T|
u_\mu(t)|dt\right)^2\le
C_1\E\left(\int_0^T|\G(t)^{-1}||\mu(t)|dt\right)^2\\ \le
C_2\sup_{t\in[0,T]}\E|\mu(t)|^2
\int_0^Tg(t)^{-1}dt<+\infty.\label{umu1}\eaaa In addition, \baaa
\E\int_0^Tg(t) |u_\mu(t)|^2dt\le
C_3\E\int_0^Tg(t)|\G(t)^{-1}\mu(t)|^2dt\bline \le
C_4\E\int_0^Tg(t)^{-1}|\mu(t)|^2dt\nonumber\\ \le
C_4\sup_{t\in[0,T]}\E|\mu(t)|^2 \int_0^Tg(t)^{-1}dt<+\infty.
\label{umu2}\eaaa Here $C_i>0$ are constants defined by $A,b,n$, and
$T$. Hence $u_\mu\in U$.
\par
 Clearly, the function $L(u,\mu)$ is strictly concave in $u$,
 and this minimization problem has
 a unique solution. Therefore, this $u=u_\mu$ is the unique solution of (\ref{min}).
\par
Further,  we consider the following problem: \baa \hbox{
Maximize}\quad L(u_{\mu},\mu)\nbline\quad\hbox{over}\quad \mu\in
L_2(\O,\F_T,\P;\R^n).\label{Lmu}\eaa
\par
 For $u=
u_\mu$, equation (\ref{solut}) gives
$$x(T)=\int_0^T e^{A(T-t)}bu_{\mu}(t)dt+e^{AT}a.$$ Hence
  \baaa &&L(u_\mu,\mu)
=\frac{1}{2}\E\int_0^Tu_{\mu}(t)^\top \G(t)
u_{\mu}(t)\,dt\bbline-\E\mu^\top\int_0^Te^{A(T-t)}b
u_{\mu}(t)dt-\E\mu^\top e^{AT}a+\E\mu^\top f. \eaaa \par

We have that
 \baaa &&\E\mu^\top\int_0^Te^{A(T-t)}bu_{\mu}(t)dt\bbline=\E\mu^\top\int_0^Te^{A(T-t)}b\G(t)^{-1}b^\top \psi(t)dt
\\&&=\E\mu^\top\int_0^Te^{A(T-t)}b\G(t)^{-1}b^\top
e^{A^\top(T-t)}\mu(t)dt\bbline
=\E\mu^\top\int_0^TQ(t)\mu(t)dt\\&&=\E\int_0^T\mu^\top
Q(t)\mu(t)dt=\E\int_0^T\E\{\mu^\top
Q(t)\mu(t)|\F_t\}dt\bbline=\E\int_0^T\mu(t)^\top Q(t)\mu(t)dt.
  \eaaa
The fifth equality here holds by Fubini's Theorem. Further, we have that
 \baaa &&\E\int_0^Tu_{\mu}(t)^\top \G(t)
u_{\mu}(t)\,dt\bbline= \E\int_0^T(\G(t)^{-1}b^\top \psi(t))^\top
\G(t)\G(t)^{-1}b^\top \psi(t)\,dt \\&&= \E\int_0^T\psi(t)^\top
b\G(t)^{-1}b^\top \psi(t)\,dt\bbline =
\E\int_0^T(e^{A^\top(T-t)}\mu(t))^\top b\G(t)^{-1}b^\top
e^{A^\top(T-t)}\mu(t)\,dt\\&& = \E\int_0^T\mu(t)^\top e^{A(T-t)}
b\G(t)^{-1}b^\top e^{A^\top(T-t)}\mu(t)\,dt \bbline=
\E\int_0^T\mu(t)^\top Q(t)\mu(t)\,dt. \eaaa
 It follows that \baaa L(u_\mu,\mu)=\E
\mu^\top (f-e^{AT}a) -\frac{1}{2}\E\int_0^{T}\mu(t)^\top Q(t)
\mu(t)\,dt. \eaaa By the Martingale Representation Theorem, there
exists $k_\mu\in L_{2,2}^{n\times d}$  such that \baaa
\mu=\oo\mu+\int_0^Tk_\mu(t)dw(t),\label{Clarc} \eaaa where
$\oo\mu\defi \E\mu$. It follows that \baaa
\E\int_0^{T}\mu(t)^\top Q(t) \mu(t)\,dt\bline
=\E\int_0^{T}\left(\oo\mu+\int_0^tk_\mu(s)dw(s)\right)^\top Q(t) \bline\biggl(\oo\mu+\int_0^tk_\mu(s)dw(s)\biggr)\,dt\\
=\int_0^{T}\oo\mu^\top
Q(t)\oo\mu\,dt\bline+\E\int_0^{T}\left(\int_0^tk_\mu(s)dw(s)\right)^\top
Q(t) \left(\int_0^tk_\mu(s)dw(s)\right)\,dt
\\
=\oo\mu^\top \left(\int_0^{T} Q(t)\,dt\right)\oo\mu\bline+\int_0^{T}\E\left(\int_0^tk_\mu(s)dw(s)\right)^\top Q(t) \left(\int_0^tk_\mu(s)dw(s)\right)\,dt\\
=\oo\mu^\top R(0)\oo\mu+\int_0^Tdt\,\E\int_0^tk_\mu(s)^\top Q(t)
k_\mu(s)\,ds \bline=\oo\mu^\top
R(0)\oo\mu+\E\int_0^Tds\int_s^Tk_\mu(s)^\top Q(t)
k_\mu(s)\,dt\\
=\oo\mu^\top R(0)\oo\mu+\E\int_0^Tk_\mu(s)^\top R(s) k_\mu(s)\,ds.
\eaaa We have used Fubini's Theorem again to change the order of
integration. Similarly, \baaa\E \mu^\top f=\oo\mu^\top \oo f+
\E\int_0^Tk_\mu(t)^\top k_f(t)\,dt, \eaaa and $\E\mu^\top
e^{AT}a=\oo\mu^\top e^{AT}a$. It follows that
 \baaa
L(u_\mu,\mu) =\oo\mu^\top (\oo f-e^{AT}a) -\frac{1}{2}\oo\mu^\top
R(0) \oo\mu\, \bline-\frac{1}{2}\E\int_0^Tk_\mu(\tau)^\top R(\tau)
k_\mu(\tau)\,d\tau\bline +\E\int_0^Tk_\mu(t)^\top k_f(t)\,dt.
 \eaaa
 Clearly, the maximum of this quadratic form is achieved for
\baaa \oo\mu=R(0)^{-1}(\oo f-e^{AT}a),\qquad \w
k_\mu(t)=R(t)^{-1}k_f(t). \label{kmu} \eaaa This means that the
  optimal solution $\w\mu$ of
 problem (\ref{Lmu}) is
 \baaa
\w\mu=R(0)^{-1}(\oo f-e^{AT}a)+\int_0^T\w k_\mu(t)dw(t). \eaaa
\par
Let $\w u(t)$ and $\w \mu(t)$  be defined by (\ref{psi})-(\ref{SMP})
for $\mu=\w\mu$, i.e., $\w u=u_{\w \mu}$.   By Lemma 1 from
\cite{D2013}, it follows that $\w k_\mu(\cdot)\in L_{2,2}^{n\times
d}$. It follows that \baaa \E\int_0^T|\w
k_\mu(t)|^2dt<+\infty.\label{final} \eaaa  It follows that
$\sup_{t\in[0,T]}\E|\w\mu(t)|^2<+\infty$.

 \par
 We found that  $\sup_\mu \inf_u L(u,\mu)$ is achieved for $(\w u,\w\mu)$.
  We have that $L(u,\mu)$ is strictly convex in $u\in U$ and
 affine in $\mu\in L_2(\O,\F,\P,\R^n)$. In addition, $L(u,\mu)$ is continuous in $u\in L_{2,2}^{n\times 1}$ given $\mu\in L_2(\O,\F,\P,\R^n)$, and
 $L(u,\mu)$
 is continuous in $\mu\in L_2(\O,\F,\P,\R^n)$ given $u\in U$.
By Proposition 2.3 from \cite{Ekland},\index{ Ekland and Temam  (1999),} Chapter VI, p.
175, it follows that
 \baa
 \inf_{u\in U}\sup_\mu L(u,\mu)= \sup_\mu \inf_{u\in U} L(u,\mu).
\label{infsup} \eaa Therefore, $(\w u,\w\mu)$  is the unique saddle point
for (\ref{infsup}).
\par
  Let  $U_f$ be the set of all $u(\cdot)\in U$ such that (\ref{sys}) holds.  It is easy to
see that \baaa\inf_{u\in U_f} \frac{1}{2}\E\int_0^Tu(t)^\top
\G(t)u(t)\,dt\bline =\inf_{u\in U}\sup_\mu
  L(u,\mu),\label{infsup1}\eaaa
 and any solution  $(u,\mu)$  of
 (\ref{infsup}) is such that $u\in U_f$. It follows that
 $\w u\in U_f$  and it is the optimal solution for problem
 (\ref{optim4})-(\ref{sys}).
  Then the proof of Theorem \ref{ThM} follows. $\Box$

\end{document}